 \documentclass[apjl]{emulateapj}

\usepackage{graphicx,amsmath,amssymb}

\shorttitle{Swift Identification of Dark Gamma-Ray Bursts}
\shortauthors{Jakobsson et al.}

\begin{document}


\title{SWIFT IDENTIFICATION OF DARK GAMMA-RAY BURSTS}

\author{P.~Jakobsson,\altaffilmark{1,2} J.~Hjorth,\altaffilmark{1} 
J.~P.~U.~Fynbo,\altaffilmark{1} D.~Watson,\altaffilmark{1}
K.~Pedersen,\altaffilmark{1} G.~Bj\"ornsson,\altaffilmark{2} and
J.~Gorosabel\altaffilmark{3,4}}

\altaffiltext{1}{Niels Bohr Institute, Astronomical Observatory, 
University of Copenhagen, Juliane Maries Vej 30, 
2100 Copenhagen, Denmark}

\altaffiltext{2}{Science Institute, University of Iceland, Dunhaga 3,
  107 Reykjav\'{\i}k, Iceland}

\altaffiltext{3}{IAA-CSIC, P.O. Box 03004, E-18080 Granada, Spain}

\altaffiltext{4}{Space Telescope Science Institute, 3700 San Martin
  Drive, Baltimore, MD 21218, USA}

\begin{abstract}
We present an optical flux vs.\ X-ray flux diagram for all known gamma-ray 
bursts (GRBs) for which an X-ray afterglow has been detected. We propose 
an operational definition of dark bursts as those bursts that are optically 
subluminous with respect to the fireball model, i.e., which have an 
optical-to-X-ray spectral index $\beta_{\mathrm{OX}} < 0.5$. Out of a sample 
of 52 GRBs we identify 5 dark bursts. The definition and diagram serve as a 
simple and quick diagnostic tool for identifying dark GRBs based on limited 
information, particularly useful for early and objective identification of 
dark GRBs observed with the \emph{Swift} satellite.
\end{abstract}

\keywords{dust, extinction --- galaxies: high-redshift --- gamma rays: bursts}

%------------------------Intro---------------------------------------
\section{INTRODUCTION}
Dark gamma-ray bursts (GRBs) remain one of the unresolved issues in GRB 
research. Shortly after the localization of the first GRB afterglows it 
became clear that not all GRBs were accompanied by detections of optical 
afterglows (OAs). In fact, a fairly large fraction, about 60--70$\%$ of 
well localized GRBs did not lead to detections at optical wavelengths 
\citep{johanDARK,lazzati02}. 
%These dark bursts were localized with 
%accuracies better than arcminutes, through a detection of either their 
%X-ray or radio afterglows. 
\par
Various scenarios have been suggested in order to shed light on 
dark bursts. The \emph{obscuration scenario} \citep[e.g.,][]{groot,taylor} 
ascribes the failed OA detection to extinction. Although there 
is evidence from X-rays \citep{gw} and damped Ly$\alpha$ absorbers 
\citep[e.g., Fig.~4 in][]{paul} of high column density of gas around many 
GRBs, the early high energy radiation from them and their afterglows can 
destroy the dust in their environment within a radius up to a few tens of 
parsecs \citep{waxman,andy,perna}. This would pave the way for the 
afterglow light, but dust in the host galaxy at larger distances could still
lead to failure in detecting the OA. In the \emph{high-redshift scenario}, 
as some fraction of bursts will be located beyond $z \gtrsim 5$ 
\citep[e.g.,][]{totani, ralph, lamb00}, the UV band, which is strongly 
affected by absorption in the Ly$\alpha$ forest, is redshifted into the 
optical band. 
%This scenario would 
%naturally explain the properties of the so-called X-ray flashes (XRFs). Here 
%the peak of the gamma-ray spectrum would be redshifted 
%into the X-ray band. This reasoning, though,  has been severely dented 
%with the detection of two XRF OAs (XRF 020903: \citet{ricker}; XRF 030723: 
%\citet{fox}). In addition, XRF 020903 has been associated with a probable 
%host galaxy at $z=0.251$ \citep{soderberg}, and \citet{XRF} have put a 
%firm upper limit of $z \lesssim 2.3$ on 
%the redshift of XRF 030723 from the afterglow spectrum.
Finally, optical faintness can arise if the OA is intrinsically dark as may 
happen, e.g., if a relativistic ejecta is decelerated in a 
\emph{low-density ambient medium} \citep[e.g.,][]{sari,taylor00}. 
\par
The dark burst fraction places important constraints on the fraction 
of obscured star formation in the Uni\-verse 
\citep{djorgovski, enrico} and the structure of star-forming 
regions \citep{lamb01, reichart}.
Statistical samples studied up to now are unfortunately quite 
heterogeneous due to large differences in localisation accuracies,
localisation time since the onset of the burst, and search strategies.
Moreover, effects of observing conditions (e.g., lunar phase) have 
generally not been taken into account in statistical studies. In many cases, 
GRBs have been considered dark if no OA was detected, irrespective of how 
inefficient the search was. In fact, there is no generally accepted 
criterion for when a GRB is considered dark. With the launch of the 
\emph{Swift} satellite it will be essential to have a quick diagnostic tool 
to flag dark bursts for immediate and/or detailed follow-up (including the 
near-IR bands) to ensure homogeneity of samples. In this Letter we present 
a GRB diagram of the optical flux ($F_{\mathrm{opt}}$) vs.\ the X-ray flux 
($F_{\mathrm{X}}$) and propose that those bursts which are optically 
subluminous with respect to the fireball model, i.e., which have an 
optical-to-X-ray spectral index $\beta_{\mathrm{OX}} < 0.5$, be defined 
as dark.
%--------------------------CURRENT STATUS---------------------------------
\section{CURRENT STATUS}
A popular working definition of dark bursts is to set a brightness 
limit at a given time after the GRB, e.g., $R > 23$\,mag at 1--2\,days 
\citep{djorgovski}. Such definitions are necessarily somewhat arbitrary 
but catch the notion of darkness very well in that the magnitude limits
and times correspond to typical search efforts and reaction times.
Another approach has been to invoke a physical definition,
specifically to require a dark burst to be a significantly obscured
burst. It has been argued that GRB~970828 \citep{djorgovski} and GRB~000210 
\citep{piro} were most likely dark because of optical obscuration.
\par
\citet{johanDARK} demonstrated that the majority ($\gtrsim$75$\%$) of GRBs 
for which searches for optical afterglow had been unsuccessful were 
consistent with no detection if they were similar to dim bursts like 
GRB~000630 in the optical band (see their Fig.~3). \citet{jens02} found that 
the dim GRB~980613 had similar properties, i.e., it would have been classified 
as a dark burst had it not been for the relatively deep search efforts.
The afterglow was neither strongly reddened nor at high redshift. This 
suggests that the classification of the majority of dark bursts was due to 
searches which simply were not sufficiently sensitive to detect the faint 
optical afterglows.
\par
\citet{berger} reached a similar conclusion for the dim
GRB~020124 and ascribed the faintness to rapid decay whereas 
\citet{jens03} demonstrated that the faintness was largely due to
the fairly large redshift of $z = 3.2$ (although not sufficiently
large for the burst to be dark due to Ly$\alpha$ absorption). 
Several studies of the rapidly localised HETE-2 burst GRB~021211 
arrived at a similar result: it would have been classified 
as a dark burst due to its rapid fading, but was found to be very bright 
after ten minutes. It was not strongly reddened and at a moderate 
redshift \citep{foxW, li, crew, pandey03}.
\par
In a study of all BeppoSAX bursts with Narrow Field Instruments 
follow-up, \citet[hereafter D03]{depas} found that most 
optically faint bursts are also X-ray faint. Some, however, appear even 
fainter in the optical than expected from X-rays. In a comprehensive study, 
\citet{rol} concluded that most GRBs can be fitted with standard fireball 
models. Only three were inconsistent with all models, i.e., fainter than 
the faintest optical expectation from X-rays. These were classified as dark. 
In addition, \citet{kp} have proposed that GRB~001025A, along with 
some other bursts, appear optically dark because their (X-ray) afterglow 
is faint and their synchrotron cooling break, $\nu_{\mathrm{c}}$, is 
located close to the X-ray band. 
\par
Recently, more homogeneous samples have been constructed based on 
BeppoSAX and HETE-2. \citet{stratta} find a dark burst 
fraction of 4/13 for a sample of bright BeppoSAX bursts. The better 
search conditions offered by HETE-2, in particular since the Soft X-ray 
Camera started to deliver accurate and fast localisations, have 
resulted in this fraction decreasing further to of the order of 10$\%$ 
\citep{lambDARK} as anticipated by \citet{johanDARK}. 
%--------------------------TABLE----------------------------------------
\begin{deluxetable*}{@{}llrrl|llrrl|llrrl@{}}
%\tabletypesize{\scriptsize}
%\rotate
\tablecaption{Gamma-Ray Bursts That Have an Unambiguous Detected 
X-ray Afterglow and an Optical Follow-Up. \label{fbeta.tab}}
\tablewidth{0pt}
\tablehead{
GRB &
Obs. &
$\beta_{\mathrm{OX}}$ & 
$R$(11\,h) & 
Ref. &
GRB &
Obs. &
$\beta_{\mathrm{OX}}$ & 
$R$(11\,h) & 
Ref. &
GRB & 
Obs. &
$\beta_{\mathrm{OX}}$ & 
$R$(11\,h) & 
Ref.
}

\startdata

970111 & SAX & $<0.83$ & $>$22.2 &      & 
990907 & SAX & $<0.69$ & $>$20.9 &      &  
020127 & CXO & $<1.24$ & $>$20.4 & (23) \\

970228 & SAX & 0.81    & 19.3    & (1)  & 
991014 & SAX & $<0.63$ & $>$22.4 &      &  
020322 & XMM & 0.51    & 23.3    & (24) \\

970402 & SAX & $<0.80$ & $>$21.5 &      & 
991106 & SAX & $<0.99$ & $>$20.2 &      &  
020405 & CXO & 0.75    & 18.3    & (25) \\

970508     & SAX & 0.69    & 21.1    & (2)  & 
991216     & CXO & 0.96    & 16.9    & (10) & 
020427$^*$ & CXO & $<0.87$ & $>$19.8 & (26) \\

{\bf 970828} & RXTE & $<0.05$ & $>$25.0 &      &
000115       & RXTE & $<1.06$ & $>$15.8 & (11) &
020813       & CXO  & 0.65    & 19.1    & (27) \\

971214 & SAX & 0.64    & 21.9    & (3)  & 
000210 & SAX & $<0.54$ & $>$23.1 & (12) &  
021004 & CXO & 0.93    & 18.4    & (28) \\

971227     & SAX & $<0.92$ & $>$20.3 &      & 
000214$^*$ & SAX & $<0.92$ & $>$19.5 & (13) &
030226     & CXO & 0.81    & 19.5    & (29) \\

980329 & SAX & 0.54    & 22.6    & (4)  &  
000528 & SAX & $<0.69$ & $>$22.5 &      & 
030227 & XMM & 0.62    & 21.7    & (30) \\

980519 & SAX & 1.06    & 18.8    & (5)  &  
000529 & SAX & $<1.09$ & $>$18.8 &      &  
030328 & CXO & 0.80    & 20.2    & (31) \\

980613 & SAX  & 0.69    & 22.6    & (6)  &  
000615 & SAX  & $<0.69$ & $>$23.1 & (14) &
030329 & RXTE & 0.86    & 14.7    & (32) \\

980703     & SAX & 0.71 & 20.1 & (7)  &  
000926     & SAX & 0.87 & 18.0 & (15) &
030528$^*$ & CXO & 0.63 & 21.1 & (33) \\

981226        & SAX & $<0.51$ & $>$23.1 &      &  
{\bf 001025A} & XMM & $<0.43$ & $>$24.3 & (16) &  
030723        & CXO & 1.07    & 20.9    & (34) \\

990123           & SAX & 0.65    & 19.4    & (8)  &  
{\bf 001109}$^*$ & SAX & $<0.30$ & $>$23.2 & (17) &
031203$^*$       & XMM & 0.80    & 21.0    & (35) \\

{\bf 990506} & RXTE & $<0.06$ & $>$23.2 &      & 
010214       & SAX  & $<0.63$ & $>$22.7 & (18) &
040106       & XMM  & 0.59    & 21.8    & (36) \\

990510     & SAX & 0.86    & 18.1    & (9)  &  
010220     & XMM & $<0.94$ & $>$21.4 & (19) &
040223$^*$ & XMM & $<0.78$ & $>$21.4 & (37) \\

990627 & SAX & $<1.02$ & $>$20.1 &      &  
010222 & SAX & 0.64    & 19.2    & (20) &
040701 & CXO & $<1.17$ & $>$18.1 & (38) \\

{\bf 990704} & SAX & $<0.43$ & $>$23.4 &      & 
011030       & CXO & $<0.59$ & $>$21.7 & (21) &
       &     &         &         &      \\

990806 & SAX & $<0.51$ & $>$23.3 &      &
011211 & XMM & 0.98    & 20.1    & (22) &
       &     &         &         &      \\

\enddata

\tablerefs{(1) \citet{gal}; (2) \citet{holger}; (3) \citet{d};
(4) \citet{reich}; (5) \citet{jaun}; (6) \citet{jens02};
(7) \citet{hol}; (8) \citet{ct}; (9) \citet{har};
(10) \citet{halp}; (11) \citet{javier}; (12) \citet{piro}; 
(13) \citet{rhoads}; (14) \citet{mai}; 
(15) \citet{johan}; (16) \citet{kp}; 
(17) \citet{jose}; (18) \citet{rol01}; (19) \citet{berg}; 
(20) \citet{stanek}; (21) \citet{vegur}; (22) \citet{palli}; 
(23) \citet{lamb02}; (24) \citet{bloom}; (25) \citet{bersier}; 
(26) This work; (27) \citet{urata}; (28) \citet{holland}; 
(29) \citet{pandey}; (30) \citet{c-t}; (31) \citet{bur}; 
(32) \citet{lip}; (33) \citet{rau}; (34) \citet{XRF}; 
(35) \citet{m}; (36) \citet{mas}; (37) \citet{sim}; 
(38) \citet{ugarte}
}

\tablecomments{A burst is marked with an asterisk if the follow-up was 
not carried out in the $R$-band, or a deeper limit was available in 
another band. In these cases, we assumed a spectral index of 0.6 to 
transform to the $R$-band. If a burst fulfills our dark burst criteria, 
i.e., has $\beta_{\mathrm{OX}} < 0.5$, its name is written in boldface. A 
total of five bursts are classified as dark according to our proposed scheme. 
The references refer to the optical follow-up; if void they are extrapolated 
from the $R$-band magnitudes listed in \citet{johanDARK}. The magnitudes 
have been corrected for Galactic extinction. \vspace{5.5mm}}

\end{deluxetable*}
%-------------------------END TABLE---------------------------------------
%-----------------------THE DIAGRAM-----------------------------------
\section{THE OPTICAL FLUX VS.\ THE X-RAY FLUX DIAGRAM}
Previous working definitions have been motivated by what makes a burst
dark: its faintness. However, in view of the results that a faint burst 
does not by itself belong to a separate class (notably GRBs~980613, 000630, 
020124, and 021211) and the study of D03 that some bursts may be optically 
faint simply because they are intrinsically faint, it is clear that 
another parameter must be invoked. D03 used the ratio of optical to X-ray
flux. Here we will use the optical-to-X-ray spectral index which is more 
directly related to physical properties of afterglows.
\par
In the simplest fireball models, which have been successfully used to 
interpret the observed properties of GRB afterglows, the spectral 
index, $\beta$ ($F_{\nu} \propto \nu^{- \beta}$), is governed by the
energy distribution of electrons, $p$, and the location of 
$\nu_{\mathrm{c}}$ \citep[e.g.,][]{sari}:
\begin{displaymath}
\beta = \left\{ \begin{array}{lllr}
   (p-1)/2 & & & \nu < \nu_{\mathrm{c}} \\
   p/2     & & & \nu > \nu_{\mathrm{c}}  
   \end{array} \right. \,.
\end{displaymath}
This result is independent of whether the outflow is collimated or
not, or whether the expansion takes place in a constant density or
stellar wind environment. In GRB afterglows, the cooling break is
frequently found to be located between the optical
($\sim$10$^{14}$\,Hz) and X-ray ($\sim$10$^{18}$\,Hz) regimes giving
rise to a break in the spectral distribution somewhere between these
two frequencies. In some cases, though, it is either positioned below 
the optical or above the X-rays.
\par
The value of $p$ is usually found to be larger than 2 \citep[$p < 2$ is 
not ruled out but requires a high-energy cutoff in the electron energy 
distribution, see e.g.,][]{dai} and smaller than 2.5. In a study of 36 
BeppoSAX X-ray afterglows, \citet{piro-p} inferred an average value of 
$p = 2.26$. In this simple picture, the average $\beta_{\mathrm{OX}}$ 
(where the subscript 'O' stands for 'optical' and 'X' for 'X-ray') is
expected to lie between 0.5 ($p=2$, $\nu_{\mathrm{c}} > 10^{18}$\,Hz)
and 1.25 ($p=2.5$, $\nu_{\mathrm{c}} < 10^{14}$\,Hz). 
%A similar range in 
%$\beta_{\mathrm{OX}}$ is predicted in the cannonball model 
%\citep[e.g.,][]{cannon}, making the 
%$F_{\mathrm{opt}}$--$\beta_{\mathrm{OX}}$ diagram (see below)
%a useful dark burst classifier, independent of the preferred afterglow 
%model. 
%For a given X-ray flux a small spectral index would give rise to 
%an optically faint burst.
\par
In a plot of $F_{\mathrm{opt}}$ vs.\ $F_{\mathrm{X}}$, optically 
subluminous bursts, i.e., bursts fainter than expected from the fireball 
model, will be situated below the line of constant $\beta_{\mathrm{OX}} = 
0.5$. In Fig.~\ref{fbeta.fig} we plot the 
$F_{\mathrm{opt}}$--$F_{\mathrm{X}}$ diagram for all known GRBs 
which have an X-ray detection and an optical detection or upper limit 
(as of August 2004). All data have been interpolated/extrapolated to 
11\,h (following D03), and are listed in Table~\ref{fbeta.tab}. For the 
upper limits we have assumed a decay index of $\alpha = 1$ 
($F_\nu \propto t^{- \alpha}$) in the extrapolation. 
We note that the significance level of reported upper limits vary
between bursts, ranging between 2$\sigma$ and 5$\sigma$.
\par
All $R$-band magnitudes in Table~\ref{fbeta.tab} have been 
corrected for foreground (Galactic) extinction using the reddening 
maps of \citet{schlegel}. At 11\,h, the optical afterglow is usually 
sufficiently bright that the host galaxy contribution is negligible, but 
whenever possible we have used the host-subtracted magnitudes reported 
in the literature. For the BeppoSAX bursts we have taken the 
1.6--10.0\,keV X-ray flux at 11\,h from D03 and calculated the flux 
density at 3\,keV. The same procedure was carried out for the
Rossi X-ray Timing Explorer (RXTE) bursts, except the 2--10\,keV X-ray 
flux was obtained from various sources in the literature. For the 
XMM-Newton and Chandra (CXO) data the flux density at 3\,keV was derived 
from the best-fit single power-law with Galactic absorption to the 
2--10\,keV data. This energy (3\,keV) was chosen as it is relatively 
insensitive to absorption and requires very little extrapolation of the 
data since it is close to the center of the bandpass with respect to 
total counts, thus yielding a reliable flux density. Data from XMM-Newton 
were reduced in a standard way using the XMM-Newton Science Analysis 
System (SAS, version 6.0.0) and the latest calibration files. The 
CXO data were reduced in a standard way using the Chandra Interactive 
Analysis of Observations (CIAO, version 3.0.2) and the latest calibration 
files (CALDB, version 2.27).
%-----------------------DISCUSSION-----------------------------
\section{DISCUSSION}
\label{dis.sec}
\begin{figure*}[!t]
\epsscale{0.9}
\plotone{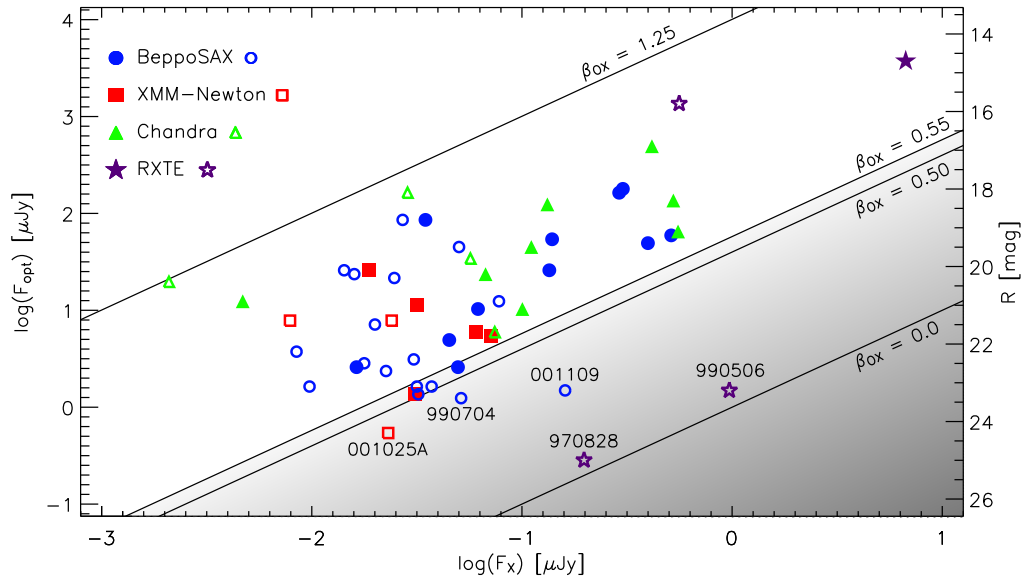}
\caption{A diagram of optical flux vs.\ X-ray flux for all bursts 
   in Table~\ref{fbeta.tab}. Optical fluxes, the corresponding
   $R$-band magnitudes shown on the right hand ordinate, and X-ray fluxes 
   have been interpolated/extrapolated to 11\,h. The magnitudes have been 
   corrected for Galactic extinction. Filled symbols indicate optical 
   detections while open symbols are upper limits. Lines of constant 
   $\beta_{\mathrm{OX}}$ are shown along with the corresponding value. We 
   define dark bursts as those which have $\beta_{\mathrm{OX}} < 0.5$.}
\label{fbeta.fig}
\end{figure*}

Bursts that fulfill our criterion $\beta_{\mathrm{OX}} < 0.5$ are 
classified as dark and are printed in boldface in
Table~\ref{fbeta.tab}. We find 5 certain dark bursts out of a 
sample of 52 GRBs, consistent with the trend that the dark burst 
fraction is approaching a level of about 10\% \citep[e.g.,][]{lambDARK}. 
%In principle, bursts with
%optical upper limits are consistent with being dark, but the limits are
%not strong enough to resolve conclusively if they are dark or simply
%dim \citep{johanDARK}. 
It is clear from Fig.~\ref{fbeta.fig} that bursts 
with no optical counterparts tend to be X-ray faint, as concluded by D03.
\par
As long as a GRB optical and X-ray flux is estimated at the same point 
in time, the burst can be located in the $F_{\mathrm{opt}}$--$F_{\mathrm{X}}$ 
diagram. To the extent that the simple external shock fireball model can be 
applied,\footnote{Assuming an unchanged OA spectrum, and that the effect of
the reverse shock does not dominate the optical flux \citep{piran}.} a 
burst will either move along constant $\beta_{\mathrm{OX}}$ lines with time 
(if the optical and X-ray bands are positioned on the same power-law
segment) or along lines with a slightly different slope 
(if $10^{14}\,\mathrm{Hz} < \nu_{\mathrm{c}} < 10^{18}\,\mathrm{Hz}$). In
the \emph{Swift} era the data will be obtained within the first hour; hence
information from the early X-ray light curve or spectrum could be used
to estimate $p$, making it possible to set a limit on $\beta_{\mathrm{OX}}$
for individual bursts (making a universal $\beta_{\mathrm{OX}}$ cutoff 
unnecessary). However, this relies on instant availability of data and is
potentially hampered by, e.g., reverse shocks and light curve fluctuations.
\par
%------------------------FIGURE-----------------------------------
%--------------------------END FIGURE----------------------------------
Dark bursts, i.e., bursts located below the line of constant 
$\beta_{\mathrm{OX}} = 0.5$ in the $F_{\mathrm{opt}}$--$F_{\mathrm{X}}$ 
diagram, are guaranteed to be special in the sense that, with respect to 
the fireball model predictions, they either have a diminished optical flux 
or an excessive X-ray flux. The former could be due to high redshift or 
obscuration, while the latter could be caused by X-ray emission lines 
\citep[e.g.,][]{reeves} or thermal emission. An 
X-ray faint burst with a low value of $p<2$ will also be classified as dark 
in this scheme. It is important to note that, using this definition of dark 
bursts, there is no assurance that we will catch \emph{all} obscured or 
high-$z$ bursts. If, for instance, for a particular burst $p=2.5$ and 
$\nu_{\mathrm{c}} < 10^{14}$\,Hz, it will have a high intrinsic 
$\beta_{\mathrm{OX}}$ value and there is no guarantee that high redshift
or optical obscuration will shift $\beta_{\mathrm{OX}}$ below 0.5.
Moreover, to answer the question why a specific burst is dark it must be 
modeled in detail; the $F_{\mathrm{opt}}$--$F_{\mathrm{X}}$ diagram is 
only a quick diagnostic tool.
\par
We may consider bursts with $0.50 \lesssim \beta_{\mathrm{OX}} 
\lesssim 0.55$ as potentially dark. We identify five such 
bursts, namely GRBs 980329, 981226, 990806, 000210, and 020322. 
If the value of $p$ is universal \citep[e.g.,][]{eli}, with 
\mbox{$p \approx 2.2$}, the lower limit on $\beta_{\mathrm{OX}}$ allowed in
the fireball model is closer to 0.6. This would shift the aforementioned 
five bursts into the dark burst category.
\par
The imminent launch of the multi-wavelength observatory \emph{Swift}, 
expected to detect $\sim$100 GRBs/year, offers a unique chance to construct 
a homogeneous sample with well-understood selection criteria. \emph{Swift} 
will reach an X-ray limit of $\sim$8\,mCrab at 60\,s and an optical limit of
$R \sim 22$\,mag at $\sim$300\,s \citep{gehrels}. For a \emph{Swift} burst 
with an X-ray afterglow detected above this flux limit and no detection in 
the UVOT image, the value of $\beta_{\mathrm{OX}}$ will be below 0.1. This
implies that the early (few minutes after the burst) \emph{Swift} data will 
be adequate to get a rough location of the burst in Fig.~\ref{fbeta.fig} 
and hence to initiate dedicated follow-up observations.
%-------------------------Acknowledgements------------------------------

\acknowledgments
We thank J.\ Bloom, C.\ Kouveliotou, D.\ Lazzati, E.\ Rol
and R.\ Wijers for discussions on dark GRBs over the years. We thank 
the anonymous referee for critical reading and useful comments 
on the paper. PJ and GB gratefully acknowledge support from a special 
grant from the Icelandic Research Council. KP acknowledges support from 
the Carlsberg foundation and from the Instrument 
Center for Danish Astrophysics (IDA). This work was supported by the Danish 
Natural Science Research Council (SNF). The authors acknowledge 
benefits from collaboration within the EU FP5 Research Training 
Network ``Gamma-Ray Bursts: An Enigma and a Tool".


%\clearpage

\end{document}